\pdfoutput=1

\documentclass[aps,showpacs,twocolumn,twoside]{revtex4}
\usepackage{graphicx}
\usepackage{amssymb,amsmath}

\begin{document}

\def\ket#1{|#1\rangle}
\def\bra#1{\langle#1|}
\def\scal#1#2{\langle#1|#2\rangle}
\def\matr#1#2#3{\langle#1|#2|#3\rangle}
\def\dalpha{{\dot{\alpha}}}
\def\ddalpha{{\ddot{\alpha}}}

\def\rb#1{\left(#1\right)}
\def\sb#1{\left[#1\right]}

\def\ti#1{\mathrm{#1}}													

\def\a2#1{\alpha_{#1}}
\def\aR2#1{\alpha_{#1}^{R}}
\def\aI2#1{\alpha_{#1}^{I}}
\def\p2#1{\pi_{#1}}
\def\pR2#1{\pi_{#1}^{R}}
\def\pI2#1{\pi_{#1}^{I}}
\def\da2#1{\dalpha_{#1}}
\def\dda2#1{\ddalpha_{#1}}
\def\daR2#1{\dalpha_{#1}^{R}}
\def\daI2#1{\dalpha_{#1}^{I}}
\def\Js{\mathrm{J}}
\def\poisson#1#2{\left\{#1,#2\right\}}

\def\Laverage{\langle L^{2}\rangle}
\def\Vaverage{\langle H'\rangle}
\def\Expect#1{\langle{#1}\rangle}

\def\komut#1#2{\left[{#1},{#2}\right]}

\def\abs#1{\left\lvert#1\right\rvert}
\def\abss#1{\abs{#1}^{2}}

\newcommand{\ud}{\mathrm{d}}
\newcommand{\MeV}{\mathrm{MeV}}
\newcommand{\e}{\mathrm{e}}
\newcommand{\ui}{\mathrm{i}}

\renewcommand{\Im}{\mathop{\text{Im}}}
\renewcommand{\Re}{\mathop{\text{Re}}}

\def\tproduct#1#2#3{\left[#1 \times #2\right]^{\left(#3\right)}}
\def\tproducta#1{\left[\alpha\times\alpha\right]^{\left(#1\right)}}
\def\tproductad#1{\left[\dot{\alpha}\times\dot{\alpha}\right]^{\left(#1\right)}}
\def\abs#1{\left|#1\right|}

\newcommand{\Freg}{F_{\mathrm{reg}}}
\newcommand{\freg}{f_{\mathrm{reg}}}
\newcommand{\Jmax}{J_{\mathrm{max}}}


\title{Quantum chaos in the nuclear collective model: II. Peres lattices}
\date{\today}
\author{Pavel Str{\' a}nsk{\' y}, Petr Hru{\v s}ka, Pavel Cejnar}
\affiliation{Institute of Particle and Nuclear Physics, Faculty of Mathematics and Physics, Charles University, 
             V~Hole{\v s}ovi{\v c}k{\' a}ch 2, 180\,00 Prague, Czech Republic}
\begin{abstract}
This is a continuation of our preceding paper devoted to signatures of quantum chaos in the geometric collective model of atomic nuclei.
We apply the method by Peres to study ordered and disordered patterns in quantum spectra drawn as lattices in the plane of energy vs. average of a chosen observable.
A good qualitative agreement with standard measures of chaos is manifested.
The method provides an efficient tool for studying structural changes of eigenstates across quantum spectra of general systems.
\end{abstract}

\maketitle

\section{Introduction}\label{sec:Introduction}

In the previous part of this article \cite{I} (Part~I), we have analyzed the correspondence between classical and quantum signatures of chaos in the Geometric Collective Model (GCM) \cite{Boh52} of nuclear vibrations.
Rotations were ruled out by the constraint of zero angular momentum.
The classical version of this model was previously shown \cite{Cej04} to exhibit a very complex dependence of regular and chaotic measures on control parameters and energy.

In Part~I \cite{I}, we have compared the classical measure $\freg$, a regular fraction of the phase space volume, with the adjunct $(1-\omega)$ of the Brody parameter.
The analysis of spectra was performed in a wide energy domain and for several values of the control parameter $B$ of the GCM Hamiltonian. 
Spectra obtained via different quantization schemes of the classical model were considered, which led to the use of three different sets of quantum levels, denoted as 5D, 2D even, and 2D odd.
Whereas the 5D spectrum corresponds to the standard five-dimensional GCM restricted to the nonrotating case, the 2D even and 2D odd spectra were derived from the quantization in the two-dimensional space of polar deformation coordinates $\beta$ and $\gamma$, with the respective condition on the parity of wave functions under the reflection of angle $\gamma$.

In all cases, the validity of the Bohigas conjecture \cite{Boh84} has been fully confirmed.
We stressed two important aspects of our calculation:
First, Bohigas' conjecture has been verified independently of the quantization scheme.
Second, because of the strong dependence of the GCM chaotic measures on energy, the competing types of level statistics have been analyzed in the local regime, i.e. for separate portions of the spectrum.
Whether the statistics in a given portion is more of the Poisson or Wigner type depends on the character (regular or chaotic, respectively) of the classical dynamics in the corresponding energy interval.
The nonmonotonous dependence of chaotic measures on energy is in contrast to a majority of systems used as case examples of chaos, in particular to all kinds of billiards (or cavities) for which the chaotic features are energy independent.

In this part of the contribution, we continue the work initiated in Part~I by considering more sophisticated techniques to describe chaos in quantum spectra.
Given that the Brody parameter captures only the short-range spectral correlations, a natural way to extend the previous results would be to consider also some measures of the long-range correlations, like e.g. the $\Delta_3$ statistics or the number variance $\Sigma^2$ \cite{rmt}.
This way we did not follow.
The reason is connected with the above-mentioned nontrivial variation of chaotic measures with energy, which would unavoidably increase statistical ambiguity of such analyses.

Instead, we employed the method invented in 1984 by Peres \cite{Per84}.
While Bohigas' conjecture, which was published in the same year \cite{Boh84}, has become a widely recognized paradigm of quantum chaos, Peres' method has been more or less forgotten.
It is mentioned in the textbook \cite{Rei92}, where some applications in integrable and non-integrable spin systems are discussed \cite{Sri90}.
An application in a billiard system was presented later \cite{Ree99}.
Today, however, Peres' name is more commonly cited in connection with his alternative definition of chaos in quantum systems \cite{Per84b}, submitted and published with a difference of just few days, which became a cornerstone for presently a quickly expanding branch of the quantum information theory \cite{Gor06}.

Nevertheless, the idea of Ref.~\cite{Per84} turns out to be very fruitful, as well.
We will show below that the method based on this idea represents a sensitive probe into the competition between regular and chaotic features in quantum spectra.
It can be applied even beyond the theory of quantum chaos, as a synoptical indicator of the changing structures across the spectrum containing possibly a very large number of states.
In the two-dimensional case, the method is graphical and may be compared to the classical method of Poincar{\' e} sections.
The spectrum of stationary states of a given quantum system with two degrees of freedom is drawn as a lattice in the plane $E\times\Expect{P}$, where $E$ is energy and $\Expect{P}$ stands for an arbitrary observable average.
This allows one to recognize ordered and disordered patterns and visually allocate regular and chaotic domains within the same energy interval.
The freedom in choosing observable $P$ makes it possible to focus on various properties of individual states and to closely follow the way how chaos sets in and proliferates in the system.

The plan of the present paper is the following:
Peres' method is introduced in Sec.~\ref{sec:PeresInvariant} and elaborated for the geometric model in Sec.~\ref{sec:Results}.
Section~\ref{sec:Discussion} shows Peres lattices calculated with two choices of operator $P$ for different values of control parameters and different quantizations.
Discussion of these results and their comparison with results of Part~I are presented.
The summary and conclusions come in Sec.~\ref{sec:Conclusions}.

\section{Peres method}\label{sec:PeresInvariant}

\begin{figure*}[t]
		\centering
		\includegraphics[width=0.67\linewidth]{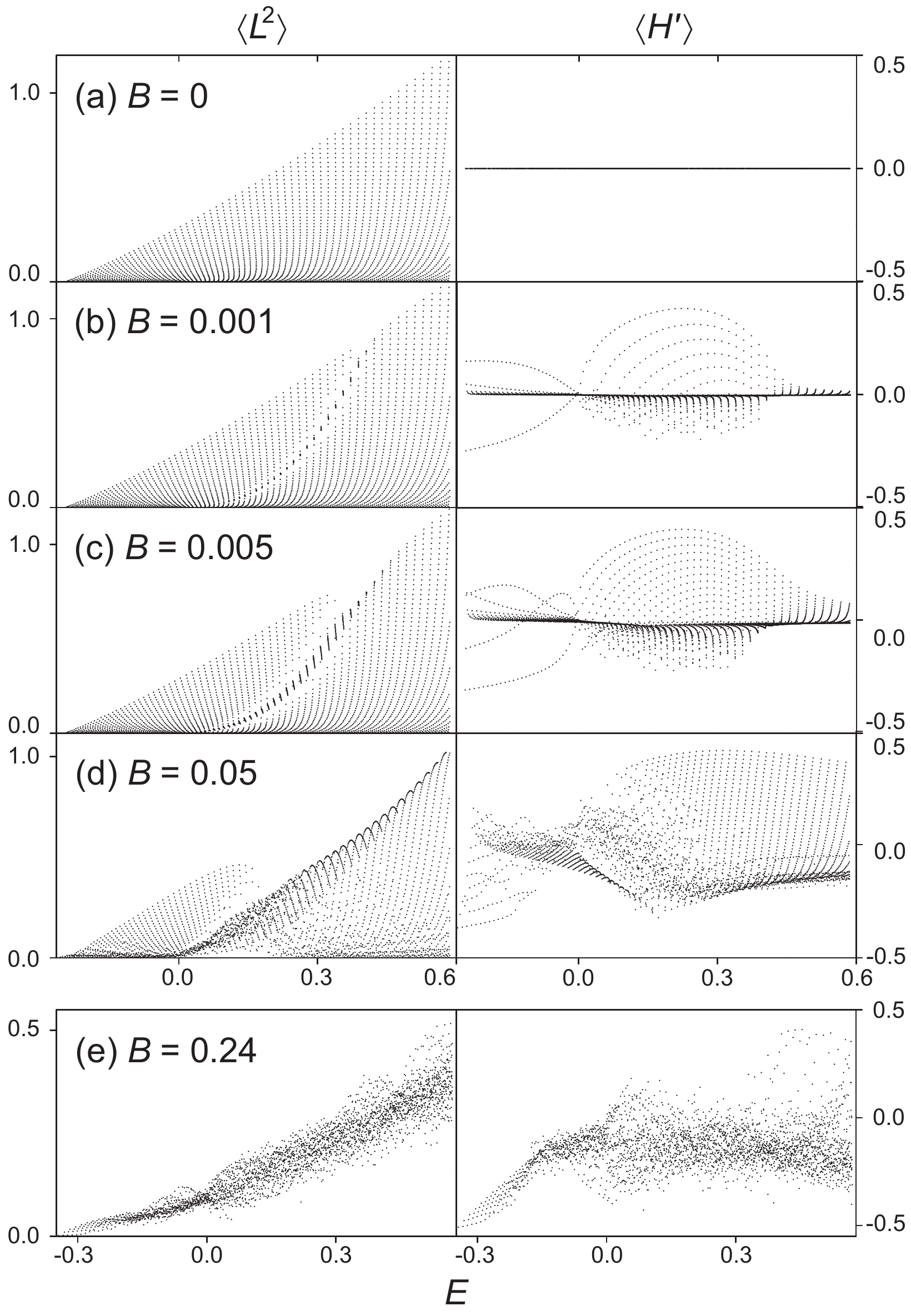}
		\caption{\protect\small 
Peres lattices for $J=0$ eigenstates of the geometric collective model in the \lq\lq 2D even\rq\rq\ quantization (see Sec.~\ref{sec:Results}). 
The points represent individual eigenstates with coordinates $E_{i}$ (energy) and $\Expect{P}_{i}$ (expectation value of the respective Peres operator).
Two Peres operators, $L^{2}$ and $H'$, are employed, the results shown in the left and right columns, respectively. 
Row (a) corresponds to the fully integrable case, $B=0$. 
Rows (b)--(d) depict the disturbance of the lattice with gradually increasing nonintegrable perturbation until reaching the most chaotic case, $B=0.24$ (e).
All quantities are given in relative units, $\hbar=5\cdot 10^{-3}$.
		}
		\label{fig:Onset}
\end{figure*}

Let us consider an {\em integrable\/} system with two degrees of freedom (a 2D system).
Apart from the Hamiltonian $H_{0}$, there must exist another integral of motion, denote it $I$, which by definition satisfies the commutation relation $\komut{H_{0}}{I}=0$.
If we plot the eigenvalues $I_{i}$ of observable $I$ against energies $E_{i}$ for individual levels (enumerated by integer $i=0,1,2,\dots$), the resulting image is formed by a lattice of regularly distributed points.
This is a straightforward consequence of the semiclassical quantization by Einstein, Brillouin, and Keller (EBK) \cite{EBK}.
An example of such a regular lattice is shown in panel (a) of Fig.~\ref{fig:Onset}, with $H_0$ and $I$ described in Sec.~\ref{sec:Results}.
In the present section we focus on the left column of the figure until specified otherwise.

The kind of lattice described above is commonly used to analyze some analytic aspects of quantum integrable systems, see e.g. Ref.~\cite{Efs04}.
In the present context, it is a natural starting point for the explanation of the Peres method. 
To continue, we introduce a perturbation $H'$ to the integrable Hamiltonian $H_{0}$,	which yields a {\em nonintegrable\/} Hamiltonian
\begin{equation}
H=H_0+\lambda H',
\label{hlina}
\end{equation}
with $\lambda$ standing for a real number measuring the strength of the perturbation.	
Obviously, $I$ does not commute with $H$ since $\komut{H'}{I}\neq0$ in general, hence $I$ is not any more an integral of motions.
Consequently, the Hamiltonian eigenstates $\ket{\psi_{i}}$ are characterized by energies $E_i$, but not by fixed values $I_i$ of observable $I$.
If we want to continue with the 2D visualization of the spectrum as introduced above for $\lambda=0$, the question raises what to draw on the vertical axis instead of $I_{i}$?
Quite naturally, Peres has chosen the expectation values $\Expect{I}_{i}=\matr{\psi_{i}}{I}{\psi_{i}}$ of observable $I$ in individual eigenstates.
This choice smoothly connects the perturbed ($\lambda\neq 0$) and the unperturbed ($\lambda=0$) cases since $\Expect{I}_{i}\to I_{i}$ for $\lambda\to 0$.

Such lattices are shown in panels (b)--(e) of Fig.~\ref{fig:Onset} (left), where the general parameter $\lambda$ from Eq.~\eqref{hlina} is replaced by the model-specific parameter $B$, see Sec.~\ref{sec:Results}.
As can be seen in panels (b)--(d), adding a small perturbation to the integrable Hamiltonian does not instantaneously break down the entire regular lattice. 
Instead, some localized seeds of distortion are created, while the rest of the lattice remains ordered in the same fashion as in the integrable case.
This scenario is in accordance with Percival's conjecture \cite{Per73} assuming that the sets of regular and chaotic eigenstates are statistically independent in the semiclassical limit $\hbar\to 0$, i.e. they do not interact with each other.
Therefore, the persisting regular parts of the lattice can be associated with surviving remnants of classical tori, while the disordered parts correspond to proliferating chaotic orbits.

As the perturbation strength $\lambda$ grows, the remnants of tori are gradually disappearing and disorder tends to increasingly plague the lattice.
This is demonstrated by an almost totally disordered lattice in panel (e) of Fig.~\ref{fig:Onset}, where only a few low lying states keeps the regular pattern.

The above-outlined visual method implies a great heuristic gain.
It allows one to judge which parts of a mixed spectrum (or, in optimal cases, which individual states) are regular and which are chaotic.
Let us stress that this is opposite to traditional methods of quantum chaos based on the spectral statistics since in that case regular and chaotic (or mixed) parts of the spectrum can only be specified by energy.
In the present approach, these parts can coexist within the same energy interval, the additional information needed for their separation being obtained from the behavior of the averages $\Expect{I}_{i}$.

In practice, there certainly exist severe limitations in the ability to distinguish from each other the regular and chaotic patterns in a finite lattice.
The identification of these patterns is further obscured by the fact that they may be superimposed on each other (as shown below).
It should be stressed that the Peres' method is {\em not\/} quantitative---it does not yield (at least not directly) a calculable measure of quantum chaos which could be compared with other measures like, e.g., the Brody parameter.

In spite of these limitations, however, the method has a great potential to disclose important features of the mechanisms governing the breakdown of integrability and rise of chaos in low-dimensional systems.
Its great advantage is that the structural information on individual eigenstates is represented by a single variable (the average of a suitably chosen observable), which allows one to use a simple visualization technique incorporating simultaneously a large number of states.
Note that in higher than 2D cases, the spectral lattice would have to be drawn in a multidimensional space, which would require to develop a sophisticated computer software for pattern recognition.
Here, as we only deal with two-dimensional systems, the most efficient software is that already built in the human brain.

Peres originally introduced his method in a more general way.
He started from the simple fact that the time average of an arbitrary classical observable is a trivial integral of motion (similarly as any function in the phase space which assigns a constant value to all points of the same trajectory).
This makes it possible, for an arbitrary system, to create an unlimited number of integrals of motion.
Of course, this does not alter the fact that the system is nonintegrable, in general.
Indeed, the functions corresponding to the new integrals are singular in the chaotic part of the phase space, hence they do not generally allow one to construct a transformation to the action-angle variables.
Nevertheless, the dependence of time averages represents an interesting probe into the system's dynamics at given energy.

The time averaging can be applied in quantum mechanics, as well.
Let us take an arbitrary Hermitian operator $P$, which in the present context will be called Peres operator.
One can construct an operator $\bar{P}$ associated with the time average of quantity $P$.
This operator has the property that the time-averaged expectation value $\Expect{P}_{\ket{\psi(0)}}=\lim_{T\to\infty}\tfrac{1}{T}\int_{0}^{T}\matr{\psi(t)}{P}{\psi(t)}dt$ for an arbitrary initial state $\ket{\psi(0)}$ can be calculated as $\Expect{P}_{\ket{\psi(0)}}=\matr{\psi(0)}{\bar{P}}{\psi(0)}$.
The operator $\bar{P}$ is readily obtained from the Heisenberg image $P_{\rm H}(t)$ of $P$ through
\begin{equation}
	\bar{P}=\lim_{T\rightarrow\infty}\frac{1}{T}\int_{0}^{T}P_{\rm H}(t)dt
	\,.
	\label{eq:TimeAverage}
\end{equation}
It is now straightforward to see that $\bar{P}$ fulfills the commutation relations $\komut{H}{\bar{P}}=0$, hence is an integral of motion.

Peres showed that in an integrable system the set of points $E_{i}$ versus $\bar{P}_{i}$ (where $\bar{P}_{i}$ is a fixed value of $\bar{P}$ in the $i$th eigenstate) forms a smoothly deformed regular lattice.
This is so irrespective of the choice of the operator $P$ used for evaluating the averages.
Even if we choose an observable $P$ which is {\em not\/} an integral of motion, $P\neq I$, the corresponding lattice for an integrable system will be ordered.
The proof makes use of the EBK quantization and the fact that in an integrable system any additional integral of motion (including $\bar{P}$) must be a function of the actions (for an integrable system, $\bar{P}$ is constant on the phase-space tori).
Therefore, any distortion of regularity of the lattice signals the onset of chaotic motions.

The expressions $\bar{P}_{i}$ and $\Expect{P}_{i}$ yield the same values and can be interchanged.
We call these values $P$-averages, while the set of points $E_{i}$ versus $\Expect{P}_{i}$ for an arbitrary (integrable or nonintegrable) Hamiltonian is denoted here as the Peres lattice. 
Note that in Refs.\cite{Rei92,Ree99}, a more pictorial term \lq\lq quantum web\rq\rq\ was proposed.

\begin{figure*}[htp]
		\includegraphics[width=0.7\linewidth]{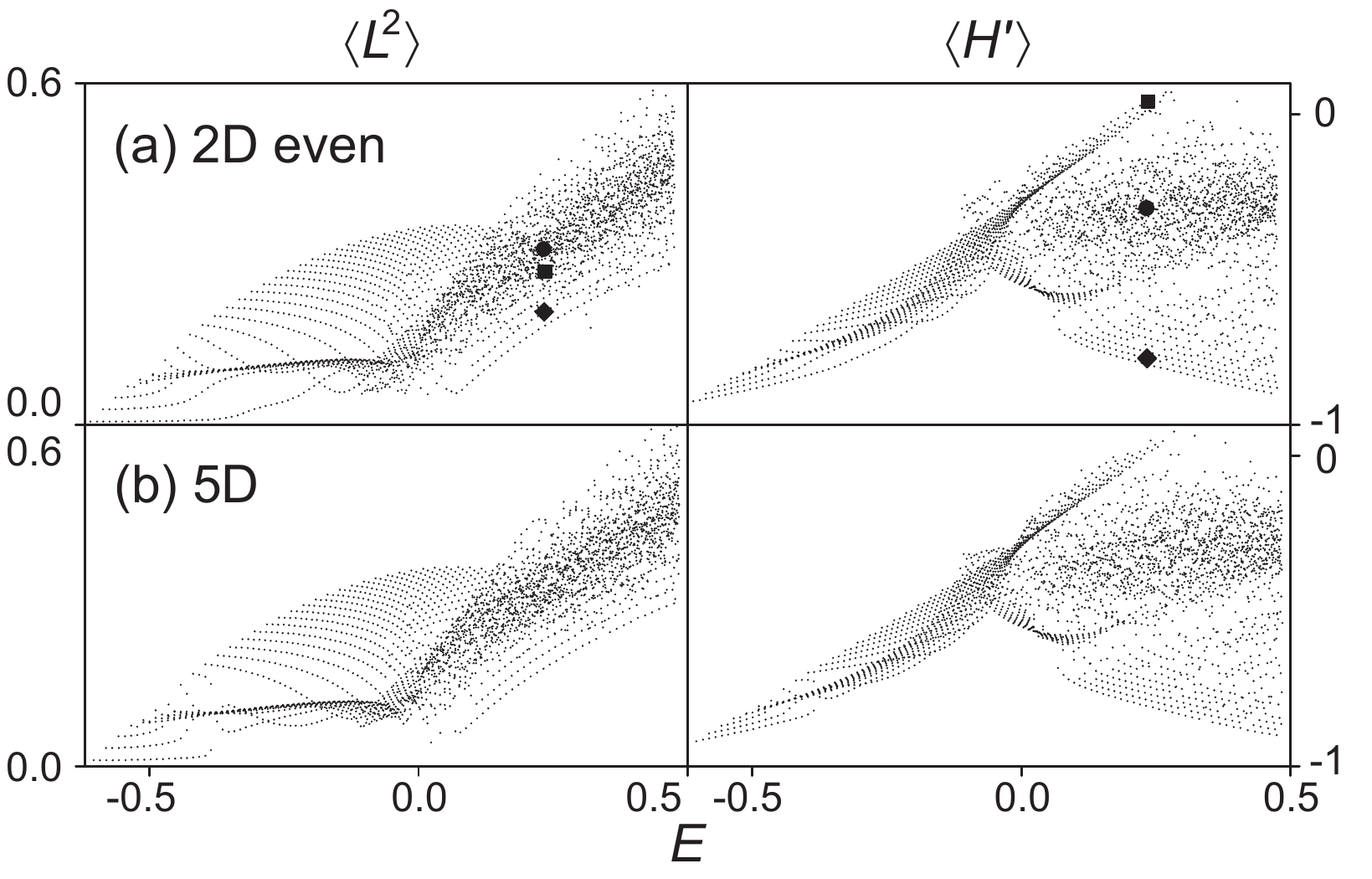}
		\caption{\protect\small 
Peres lattices of the GCM at $B=0.62$ with averages of $L^{2}$ (left) and $H'$ (right) for 2D even (row a) and 5D (row b) quantizations ($\hbar=5\cdot10^{-3}$).
In row (a), three states denoted by full symbols are identified in both lattices, demonstrating that the assignment of a given state to a
regular or chaotic part of the lattice does not depend on the choice of the Peres operator.
		}
		\label{fig:Peres062}
\end{figure*}

We want to stress that there is no restriction in the choice of the Peres operator $P$.
Different choices give different lattices, but the separation of levels into regular and chaotic parts of the lattice is independent of the choice.
This consequence of the Percival conjecture will be discussed below.
It is illustrated in Fig.~\ref{fig:Onset}, where the right-hand column shows lattices for another Peres operator than that used in the left-hand column.
The rows correspond to the same values of the perturbation strength.
We observe that the overall degree of chaos in each adjacent pair of images is about the same.
Moreover, it can be shown (cf. Fig.~\ref{fig:Peres062}) that the states allocated in the regular (chaotic) part of one lattice lie in the regular (chaotic) part of the other lattice, as well.
[Note that illusive differences in the numbers of points on both sides of Fig.~\ref{fig:Onset} (and some of the forthcoming figures) are caused by eventual accumulation of multiple points with very close coordinates.]

\section{Hamiltonian and Peres operators}\label{sec:Results}

In this section we briefly introduce the Hamiltonian of the Geometric Collective Model (GCM) of atomic nuclei \cite{Boh52} in a nonrotating regime and suitable Peres operators.
The model has been discussed in Part~I \cite{I}.
The GCM Hamiltonian $H=H_{0}+BH'$ consists of the integrable part
\begin{equation}
H_{0}=T-\beta^{2}+\beta^{4}
\end{equation}
and a nonintegrable perturbation
\begin{equation}
H'=\beta^{3}\cos{3\gamma}
\,,
\label{eq:V}
\end{equation}
where	polar coordinates $\beta$ and $\gamma$ stand for dynamical variables (shape parameters) and $T$ is the kinetic energy, which involves the associated momenta \cite{I}.
The corresponding Cartesian coordinates read as $(x,y)=(\beta\cos\gamma,\beta\sin\gamma)$, with the momenta transformed accordingly.
Parameter $B$ is the perturbation strength, a model-specific version of the above-introduced general variable $\lambda$.
This choice corresponds to the GCM potential $V=A\beta^{2}+B\beta^{3}\cos{3\gamma}+C\beta^{4}$ \cite{I} with $(A,C)=(-1,+1)$.
As $C$ is positive, the Hamiltonian for any energy $E$ describes motions confined within a finite domain of $\beta$. 
The three degenerate global minima of the potential $V$ are located at $\beta>0$, $\gamma=\tfrac{\pi}{3}$ or 0 (for $B>0$ or $<0$, respectively),  and a single local maximum is at $\beta=0$.
As in Part~I, all quantities are considered dimensionless.

We take into account two different and physically relevant quantization schemes, which are connected with two- and five-dimensional versions of the system (hereafter referred to as 2D and 5D cases, respectively) \cite{I}.
The kinetic term $T$ is different for both schemes,	nevertheless in both cases it is proportional to the squared Planck constant over $2K$, where $K$ stands for an effective mass parameter of the system.
The fraction $\kappa=\hbar^{2}/K$ is called the classicality parameter.
The value of this parameter adjusts the absolute density of quantum spectra.
In the following we set $K=1$ and vary the value of $\hbar$.
The diagonalization of the Hamiltonian is performed in the appropriate 2D or 5D harmonic oscillator bases.
The 2D case is further split to even and odd case, referring to the symmetry or antisymmetry with respect to the $\gamma\to-\gamma$ inversion. 

A set of eigenenergies and eigenvectors is obtained, for which the Peres lattice is constructed.
We consider two types of the Peres operator.
The first one is identified with the square of the angular momentum operator $L$ connected with the rotations varying angle $\gamma$.
In the 2D case, this is the Casimir invariant of the O(2) algebra of rotations in the $(\beta,\gamma)$ plane:
\begin{equation}
L^2_{\rm 2D}=\hbar^2\frac{\partial^2}{\partial\gamma^2}
\quad \leadsto\ \hbar^2 m^2\,,\quad m=0,3,6,\dots,
\label{L2d}
\end{equation}
cf. Eq.~(7) of Ref.~\cite{I}.
Note that the $L^2_{\rm 2D}$ eigenvalues, indicated in the last equation, involve multiples of 3 due to the required symmetry of the eigenfunctions with respect to rotations about angle $2\pi/3$ \cite{I}.
In the 5D case, $L^2$  is the Casimir invariant of the GCM algebra O(5) \cite{Cha76} restricted to value $J=0$ of the O(3) angular momentum (null rotations in the ordinary space).
We have
\begin{eqnarray}
L^2_{\rm 5D}=\frac{\hbar^2}{\sin 3\gamma}\frac{\partial}{\partial\gamma}\sin 3\gamma\frac{\partial}{\partial\gamma}
\quad  \leadsto && \hbar^2 v(v+3)\,,
\label{L5d}\\
&& \ v=0,3,6,\dots,
\nonumber
\end{eqnarray}
cf. Eq.~(5) of Ref.~\cite{I}.
The eigenvalues of this operator are enumerated by an integer $v$ (in the nuclear context called seniority), which for $J=0$ again has only the values equal to multiples of 3 \cite{Cha76}.

For the Hamiltonian eigenstates, $L^2_{\rm 2D}$ and $L^2_{\rm 5D}$ take the values from Eqs.~\eqref{L2d} and \eqref{L5d}, respectively, only for $B=0$, in which case the $L^2$ operators commute with the Hamiltonian.
For $B\neq 0$, the energy eigenstates $\Psi_i(\beta,\gamma)$ are mixtures of states with different values of $m$ or $v$.
In any case, the average $\Laverage_i$ quantifies oscillations of the wave functions $\Psi_i$ in the direction of angle $\gamma$ (examples shown below).

The second Peres operator used in our analysis is identified with the perturbation $H'$ from Eq.~\eqref{eq:V}.
It is worth noting that the expectation value of $H'$ in an eigenstate $\ket{\psi_i}$ of the general Hamiltonian \eqref{hlina} coincides with the derivative $dE_i/d\lambda$ of the $i$th energy level at the given value of the control parameter.
In the present case, the angular part of the $H_0$ eigenstates has the property that $\Expect{\cos 3\gamma}_i=0$, hence
\begin{equation}
\Expect{H'}_i\bigr|_{B=0}=\frac{dE_i}{dB}\biggr|_{B=0}=0
\end{equation}
(see the upper right panel of Fig.~\ref{fig:Onset}).
The vanishing slope at $B=0$ is consistent with the symmetry of the spectrum $E_i(B)$ under the reflection $B\mapsto-B$. 
If $B\neq 0$, however, the average is generally nonzero and satisfies $\Expect{H'}_i=(E_i-\Expect{H_0}_i)/B$.
The disturbances in the $P=H'$ lattice therefore show up as departures of individual points from the line $\Expect{H'}_i=0$, as seen in the right-hand column of Fig.~\ref{fig:Onset}.
This facilitates the visual inspection of results.

\section{Results and discussion}\label{sec:Discussion}

\subsection{Comparison of Peres lattices}

\begin{figure*}[tbp]
		\includegraphics[width=0.65\linewidth]{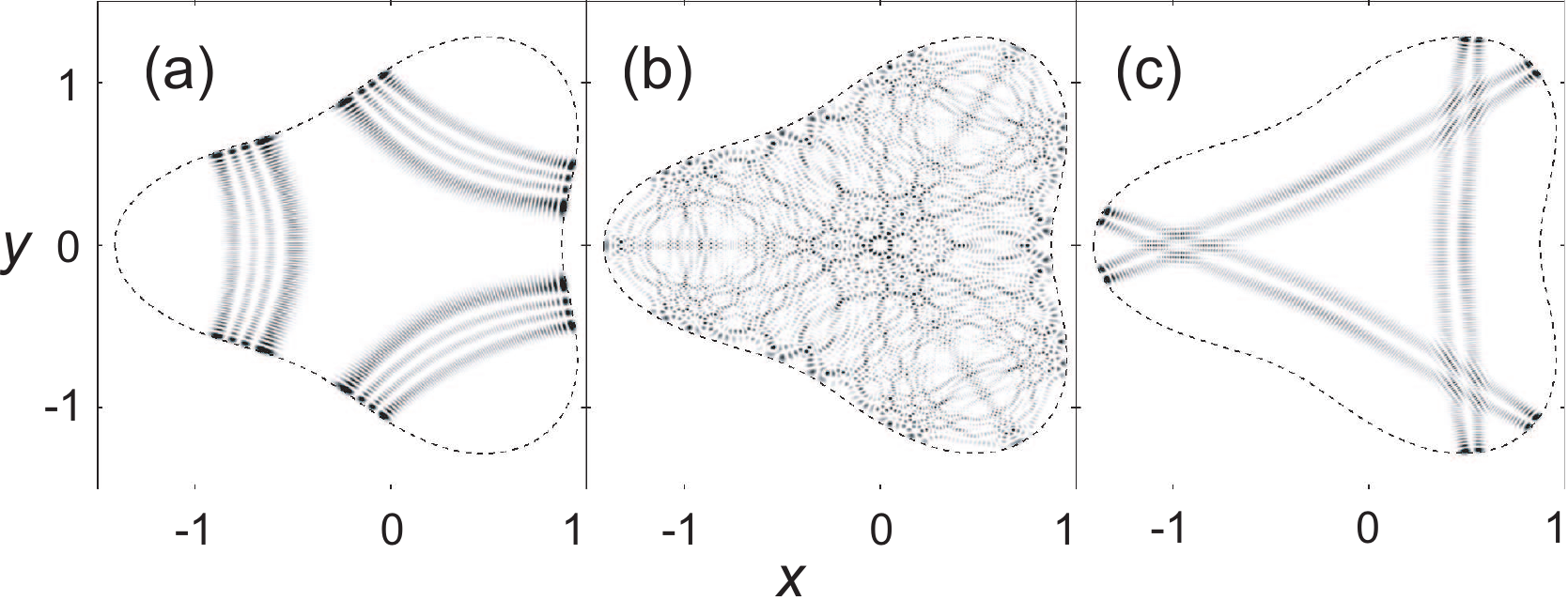}
		\caption{\protect\small
Squared wave functions for the three marked states from Fig.~\ref{fig:Peres062} (2D even quantization).
The states (a) and (c), which correspond to the square and the diamond, respectively, are taken from the regular part of the lattice (the 1995th and 1885th level, respectively) and exhibit well pronounced quantum scar effects.
The state (b), associated with the bullet, belongs to the chaotic part (the 1890th level) and shows an ergodic behavior.
		}	
		\label{fig:wf_chreg}
\end{figure*}

The effects accompanying the decay of regularity in Peres lattices were discussed from a general viewpoint in Sec.~\ref{sec:PeresInvariant}.
Figure~\ref{fig:Onset}, which corresponds to the GCM in the 2D even quantization with the above two choices of the Peres operator, illustrates the gradual transition from ordered to disordered lattices.
Note that the case $B=0.24$ depicted in row (e) corresponds to the minimum of the classical regular fraction $\freg$, see Part~I \cite{I}.
The gradual onset of chaos in these lattices will be analyzed in more details in Sec.~\ref{se:break}.

In Fig.~\ref{fig:Peres062} we compare Peres lattices obtained (for both Peres operators) in the 2D even and 5D quantizations.
The value $B=0.62$ belongs to the island of strongly pronounced regularity close to the resonance of $\beta$ and $\gamma$ vibrations, see Sec.~\ref{se:re}.
The regularity shows up as a large area of ordered points, which starts at the lowest negative energies and spreads over to positive energies, where it is joined by a rising chaotic area.
Despite the spectra for different quantizations show significant differences \cite{I}, the form of Peres lattices is rather similar.

In order to demonstrate the coincidence of the regular and chaotic regions in the lattices for different Peres operators, we have highlighted three of the states in the first row of Fig.~\ref{fig:Peres062}, marking them by a square, a bullet, and a diamond.
Probability densities for the corresponding wave functions are depicted in Fig.~\ref{fig:wf_chreg}.
The wave functions as well as the location of the respective points in the Peres lattice show that the square and the diamond correspond to regular states, while the bullet represents a chaotic state.
In Fig.~\ref{fig:Peres062}(a) we see that this assignment is consistent for both choices of the Peres operator.
Let us note that both regular levels (a) and (c) in Fig.~\ref{fig:wf_chreg} exhibit a large increase of the wave-function magnitude in a region where a certain periodic trajectory oscillates in the classical case \cite{Sto99}.
		
Although, as emphasized above, we can choose an arbitrary operator for plotting the Peres lattice, Fig.~\ref{fig:Peres062} indicates that some operators may be more suitable than others.
For some choices, a part of the regular region in the lattice can pervade into the chaotic area and hide there behind a disordered mesh of points.
In such cases, one cannot decide whether a level inside a chaotic region is indeed chaotic.
(On the other hand, overlapping regular areas form a regular area again.)
While there is no doubt that these observations demonstrate limitations of the Peres method, one can improve its resolution by employing several incompatible Peres operators.
Indeed, Fig.~\ref{fig:Peres062} (right) shows that for the three highlighted states a better choice of Peres operator is $P=H'$.

Figure~\ref{fig:Peres109} displays Peres lattices for 2D even quantization with different classicality constant $\kappa$.
In Part~I, we have shown that by tuning the value of $\kappa$ one scales the absolute density of quantum states but does not affect statistical properties of the spectra.
Here we want to show that these changes do not influence the main features of the Peres lattice.
The variations of the lattice with $\kappa=4,25,100\cdot10^{-4}$ for both Peres operators are observed in rows (a)--(c) of Fig.~\ref{fig:Peres109}.
Note that $B=1.09$ is a value of the control parameter for which the system exhibits very rich structures	with well pronounced minima and maxima of the dependence classical regular fraction $\freg$ on energy (see the insets).
A decrease of the $\kappa$ value increases the density of states (the system gets closer to the classical limit) and serves as a zoom into the sea of levels: one can see finer details of the lattice but (because of computational limits) a smaller fraction of the spectrum is available. 
For a comparison, the box in all panels of the same column encloses a fixed region of energy $\times$ P-average.
It is seen that the structures observed in the lattice become wealthier in details as $\kappa$ decreases, but the overall appearance of the relevant part of the lattice remains the same.

\subsection{Links to classical dynamics}

\begin{figure*}[tbp]
		\centering
		\includegraphics[width=0.75\linewidth]{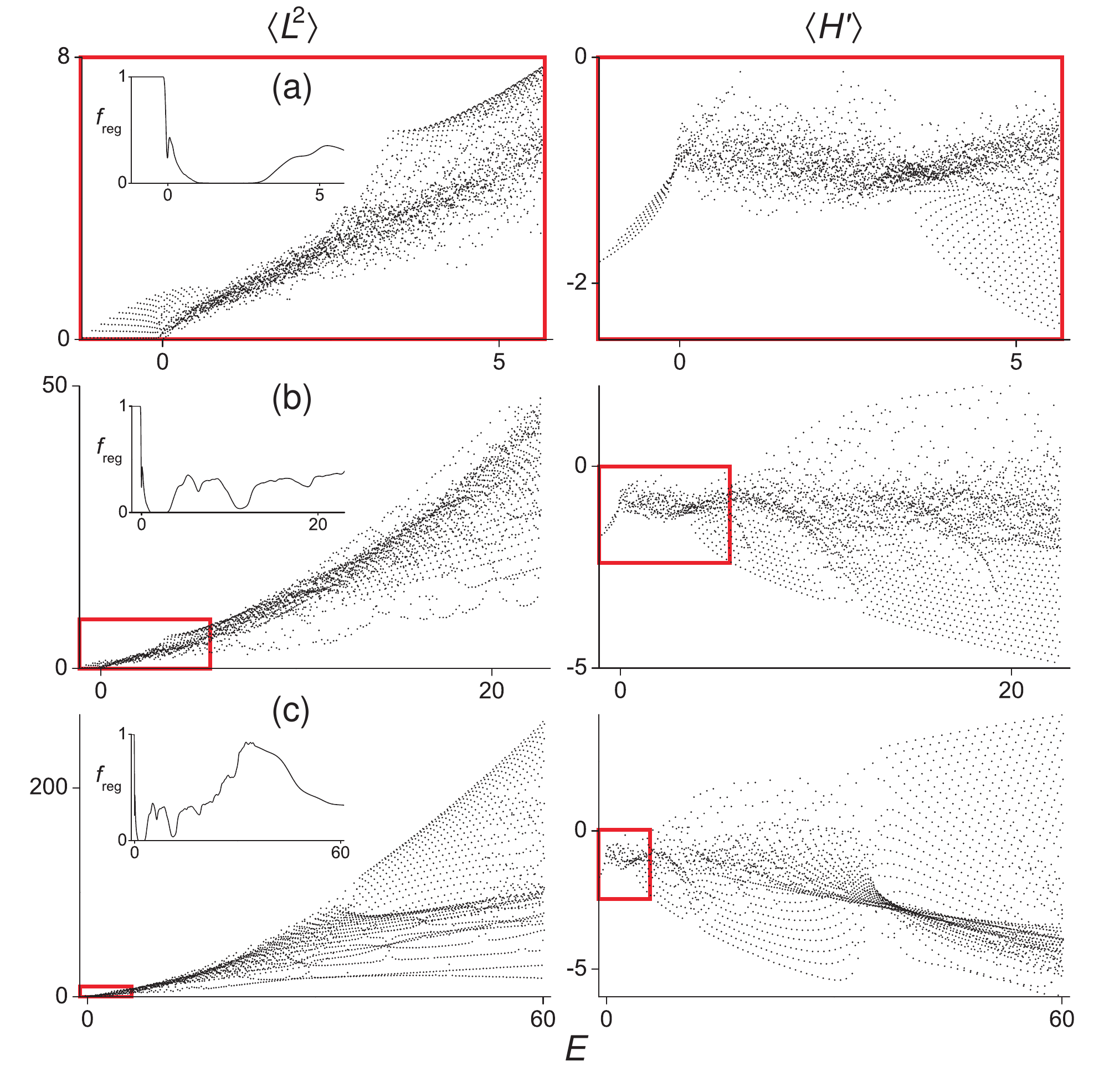}
		\caption{\protect\small 
The GCM Peres lattices for $B=1.09$ and different values of the Planck constant: $\hbar=0.02$ (a), 0.05 (b), and 0.1 (c). 
The 2D even quantization was employed with the same Peres operators as in the previous figures. 
Lower values of $\kappa$ yield denser spectra, which can therefore be evaluated only in narrower energy intervals (numerical limitations).
The box (red online) encloses the same area in the three panels of each column.
Insets show the dependence of the classical regular fraction $\freg$ on energy (see Part~I \cite{I}) for the energy domain displayed in each row.
One can observe that the $\freg$ dependences are correlated with ordered and disordered areas in the corresponding lattices in both columns.
		}		
		\label{fig:Peres109}
\end{figure*}

In Part~I, we investigated the connection between the classical measure of regularity $\freg$ (the fraction of the regular phase-space volume) and the quantum measure represented by $(1-\omega)$, where $\omega$ stands for the Brody parameter. 
Our conclusion was that both measures entail qualitatively the same energy-dependent behavior, irrespective of the method of quantization.
In the present context, new questions appear, namely: 
How strong is the correspondence between the behavior of $\freg$ and the character of Peres lattices?
Or more specifically, is there a correlation between ordered (disordered) parts of the lattice and regular (chaotic) parts of the phase space?

In order to find an answer we plot figures showing the dependence $\freg(E)$ in the insets of Fig.~\ref{fig:Peres109}.
Pure visual inspection discloses strong correlations between the increase of $\freg$ and the occurrence of regular domains in Peres lattices for both operators $\Laverage$ and $\Vaverage$.

Following the dependence in row (a) of  Fig.~\ref{fig:Peres109} we find $\freg=1$ for negative energies and observe fully regular patterns in the corresponding part of both lattices.
This is the domain where the quadratic-well approximation of the GCM potential is valid.
Note that the regular pattern at low energies is present in Peres lattices for all values of parameter $B$.
At energies just below $E=0$ the classical regularity begins to drop (forming a small fold at $E\approx0$), which is manifested in the lattice of $\Laverage$ by a band of disordered points with a small tail penetrating	to the regular area at zero energy.
Passing through the totally chaotic area with $\freg=0$ at $1\lesssim E\lesssim3$, the regularity begins to rise again.
This is accompanied by a formation of a new regular pattern in the lattices at large values of $\Laverage$ or small values of $\Vaverage$.

One can switch to rows (b) and (c) of Fig.~\ref{fig:Peres109} and continue in the same manner.
A remarkable phenomenon appears at $E\approx 35$, where $\freg$ reaches for a while the value of full regularity.
This somewhat surprising behavior was discovered in Ref.~\cite{Cej04}. 
Here we may trace the signatures of regularity in the two Peres lattices.
For instance, the lattice in the lower right panel gets locally contracted to a narrow interval of $\Vaverage$ and develops a highly organized pattern for $E>35$.
Distortions of this pattern start appearing at energies $E\gtrsim 50$, where the classical regularity is decreasing again.
Let us note that for energies much above the range shown in Fig.~\ref{fig:Peres109}, the order increases to the asymptotic value $\freg\to 1$ for all values of parameter $B$.
This is due to the $\beta^{4}$ term of the potential which becomes increasingly important at high energies, generating predominantly regular dynamics.

\begin{figure}[tbp]
		\centering
		\includegraphics[width=\linewidth]{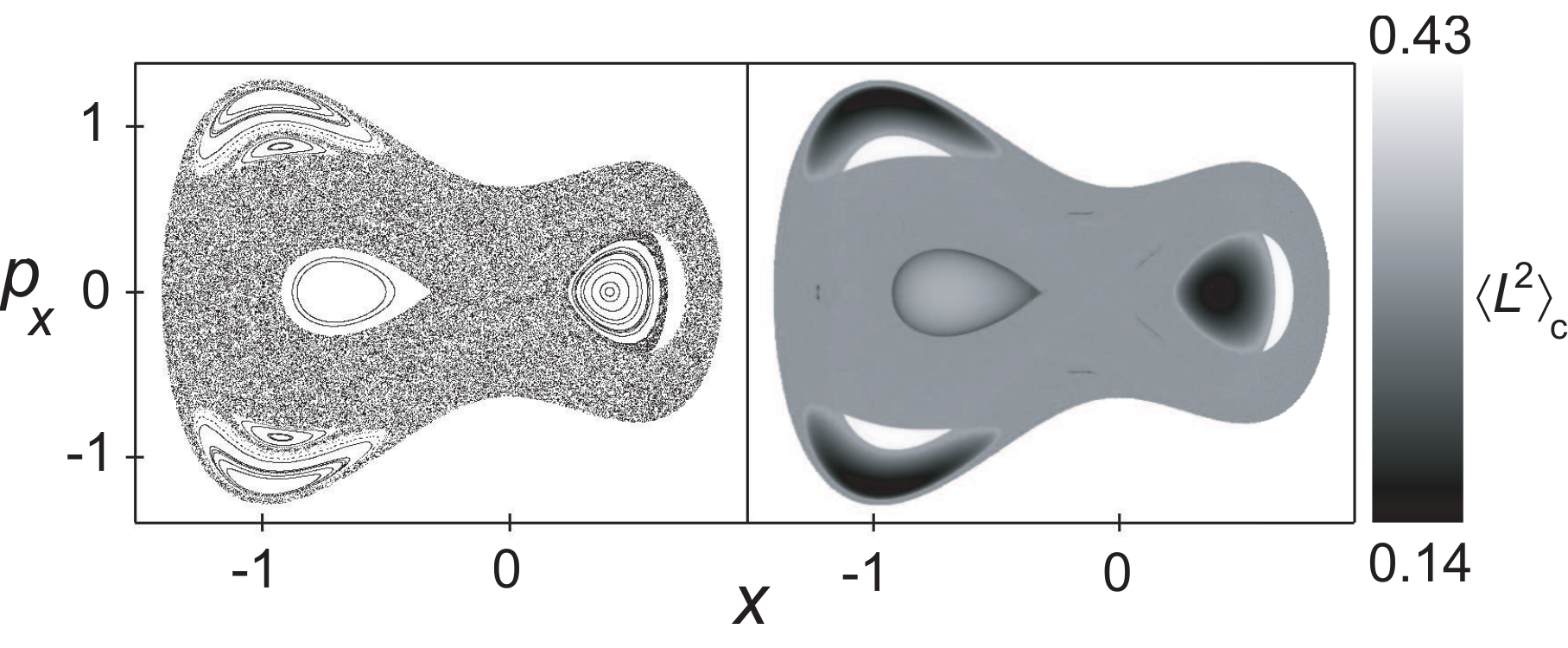}
		\caption{\protect\small 
The $y=0$ section of the classical phase space (coordinate $x$ versus momentum $p_x$) for $B=0.62$ and $E=0.2$.
Left: Poincar{\' e} section formed by crossings of $10^2$ randomly chosen trajectories with the plane of the section ($10^3$ crossings for each trajectory).
Right: Density plot of the Peres invariant $\Laverage$ calculated classically across the section (on a mesh of $500\times 500$ points).
Dark and light regions (low and high values of the Peres invariant, respectively) contain regular trajectories (see the left panel) and simultaneously correspond to regular domains in the Peres lattice [see Fig.~\ref{fig:Peres062}~(a) at $E\approx0.2$, where the regular domains are located at the lower and upper sides of the lattice].
		}
		\label{fig:Peres062c}
\end{figure}

As explained in Sec.~\ref{sec:PeresInvariant}, Peres invariants can be introduced on both classical and quantum levels.
One can determine the classical analogue of the Peres operator $P$ and calculate its average $\Expect{P}_{\rm c}$ over an arbitrary trajectory.  
In this way, a function in the classical phase space can be constructed for any Peres invariant.
In the right-hand panel of Fig.~\ref{fig:Peres062c}, the function $\Laverage_{\rm c}$ is shown (coded in shades of gray) for a certain values of the control parameter and energy on the $y=0$ section of the phase space.
Remind that $(x,y)$ represent Cartesian counterparts of the polar coordinates $(\beta,\gamma)$, while $(p_{x},p_{y})$ are the corresponding momenta, and that there is no difference between 2D and 5D cases on the classical level \cite{I}.

On the left-hand side of Fig.~\ref{fig:Peres062c}, the standard Poincar{\' e} section is plotted for 100 crossing trajectories.
We observe that the chaotic area identified in the Poincar{\' e} section is covered by one shade of gray in the map of $\Laverage_{\rm c}$.
This follows from the ergodicity of chaotic motions, which ensures that any vicinity of each point in a chaotic phase-space domain is visited by a single trajectory.
Therefore, almost the whole domain yields a single value of the classical Peres invariant (with exceptions including periodic orbits that however fill only a zero-measure subset of the phase space).
On the other hand, in the regular islands of the Poincar{\' e} section the shade of the $\Laverage_{\rm c}$ image gradually changes.

The corresponding Peres lattice was shown in Fig.~\ref{fig:Peres062}~(a), with squared wave functions of the selected states depicted in Fig.~\ref{fig:wf_chreg}.
The agreement with classical results in Fig.~\ref{fig:Peres062c} is remarkable.
The trajectory responsible for the \lq\lq scar\rq\rq\ of the wave function in Fig.~\ref{fig:wf_chreg}~(a) passes the central regular part of the phase-space section in Fig.~\ref{fig:Peres062c} close to $(x,p_{x})=(-0.8,0)$, yielding a medium value of $\Laverage_{\rm c}$.
Indeed, the respective state (denoted by the square) is localized in the medium part of the $\Laverage$ Peres lattice. 
The trajectory contributing to the wave function in Fig.~\ref{fig:wf_chreg}~(c) falls to the dark regular regions of the density plot in Fig.~\ref{fig:Peres062c}, which again corresponds to the value of $\Laverage$ for the respective point (the diamond).
A comparison of Figs.~\ref{fig:Peres062c} and \ref{fig:Peres062}~(a) indicates an excellent correspondence between the results based on the Poincar{\' e} and Peres methods (if the latter one is supplemented by a classical calculation as in the right-hand panel of Fig.~\ref{fig:Peres062c}).

\subsection{Decay of regularity}
\label{se:break}

We return now to Fig.~\ref{fig:Onset} in order to discuss in more detail the mechanism of the transition from the integrable dynamics at $B=0$ to the chaotic $B>0$ regime.
As pointed out above, in the integrable case (row a) the quantity $L^{2}$ is an integral of motion and $\Vaverage$ is identically zero.

Let us look at the curved chains of points apparent in panel (a) of Fig.~\ref{fig:Onset} (left).
These chains, which begin at $\Laverage=0$ and lead upwards, connect states with a constant sum $N=n_{\beta}+m_{\gamma}$, where $n_{\beta}$ is the radial quantum number and $m=3m_{\gamma}$ represents the angular-momentum quantum number (the latter increases towards the upper end of the chain).
If the perturbation is turned on,	some of the points within the same chain start moving against each other, forming a kind of \lq\lq condensation centers\rq\rq; see the left panels (b) and (c).
A detailed inspection discloses that the most affected levels lie in the short stretches of the chains which are nearly parallel with the vertical axis.
In other words, the perturbation is most efficient for the states which are very close in energy and in the value of $\Laverage$.
Indeed, exploiting the perturbation theory, we can say that if the perturbation matrix element is nonzero (at $B=0$, the operator $H'$ couples only the states differing in $m_{\gamma}$ by $\pm 1$), the proximity of levels leads to an increased mixing.
This in the present case shows up as an attraction to a common value of $\Laverage$ for the whole bunch.
For $B=0$ the vertical stretches are developed in the chains located within the energy interval $0\lesssim E\lesssim0.4$ and, consequently, the corresponding levels in these chains are most vulnerable if $B$ starts to increase.
This is why small perturbations affect first only a very limited part of the lattice, as observed in panels (b) and (c).

If we continue increasing the perturbation, more and more levels become influenced by the interaction. 
For sufficiently large values of $B$, the levels start	interacting between neighboring chains and the whole structure gradually breaks down, see Fig.~\ref{fig:Onset}~(d).
For $B=0.24$ (panel e) the lattice is totally disintegrated. 
We have just reached the most chaotic parameter region, where only the deepest levels form a regular lattice due to the validity of the quadratic-well approximation.

\begin{figure}[tbp]
		\centering
		\includegraphics[width=0.9\linewidth]{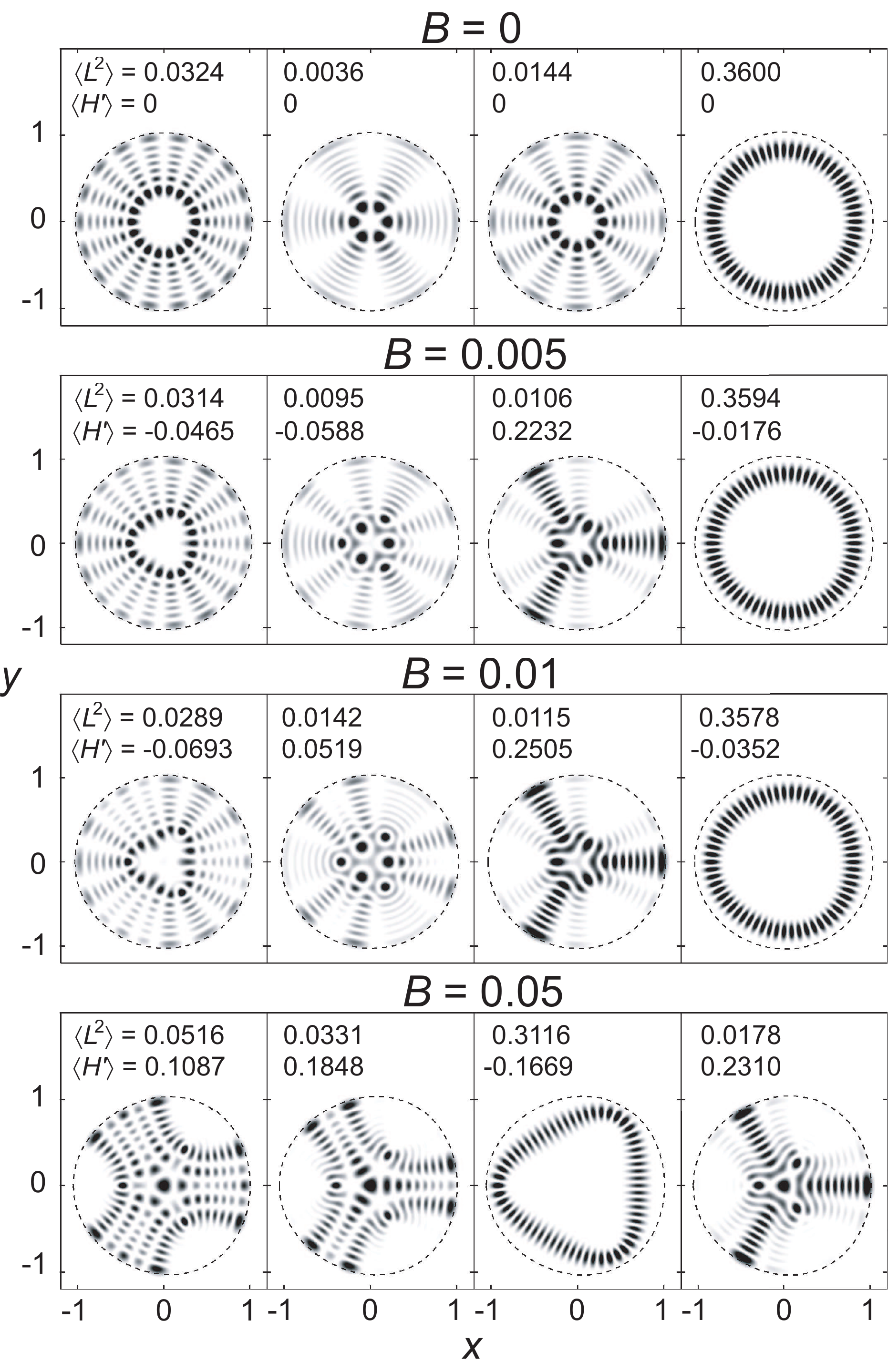}
		\caption{\protect\small 
Squared wave functions of four successive eigenstates (states no. $i=52,\dots,55$ at $E$ from 0.059 to 0.076) of the integrable $B=0$ Hamiltonian (for $\hbar=0.02$) in the 2D even quantization (the first row) and the same states for $B=0.005$, $0.01$, and $0.05$ (the second, third, and fourth rows, respectively).
The respective values of both P-averages are given in each panel.
The first three states (arranged in columns) are more sensitive to the perturbation than the fourth one, in accord with the lattices in Fig.~\ref{fig:Onset} (see the text).
		}		
		\label{fig:wf_onset}
\end{figure}

The size of the perturbation can be quantified with the aid of the other Peres operator, i.e. by the value of $\Vaverage$, which is displayed in the right-hand column of Fig.~\ref{fig:Onset}.
Rows (b) and (c) help to discover that not only the levels with $E>0$, but also also a few of those with $E<0$ become disturbed by a small perturbation (this was not visible in the left-hand panels).
For $E>0$, we observe several regular arcs of points at $\Vaverage>0$ and some more disordered points with $\Vaverage<0$.
Both these groups of points correspond to the \lq\lq condensation centers\rq\rq\ apparent in the left-hand panel.
The $\Vaverage>0$ part of the lattice contains states with $\Expect{\cos{3\gamma}}_i>0$, hence $\gamma$ centered around values 0, $\tfrac{2\pi}{3}$ and $\tfrac{4\pi}{3}$ (saddle points of the potential).
On the other hand, the $\Vaverage<0$ part collects states with $\Expect{\cos{3\gamma}}_i<0$, hence $\gamma\sim\tfrac{\pi}{3},\pi,\tfrac{5\pi}{3}$ (global minima of the potential).
Examples of both these types of wave functions will be given later in Fig.~\ref{fig:wf_onset}.
It is somewhat surprising that the more regular part of the lattice is connected with the states localized in the saddle-point regions, whereas the states localized around the minima seem to be more chaotic.

For moderate perturbation strengths, great majority of points in Fig.~\ref{fig:Onset} (right) remains located at $\Vaverage=0$, indicating the absence of structural changes.
These points correspond to the unperturbed parts of the lattice in the left-hand column.
As $B$ increases, however, both positive and negative halves of the $\Vaverage$ lattice become increasingly populated and finally the negative (irregular) part captivates absolute majority of points (row e).
This agrees with the disordered form of the lattice in the left-hand column.

The changes of the Hamiltonian eigenstates accompanying the above-described evolution of the Peres lattices are illustrated in Fig.~\ref{fig:wf_onset}.
Its first row presents four unperturbed ($B=0$) wave functions (probability distributions in the 2D even quantization), while the second, third, and fourth rows demonstrate the effects of perturbation (for $B=0.005$, $0.01$ and $0.05$, respectively) on the same states.
The states correspond to four successive energy levels, the associated values of both P-averages being given in each case. 
Note that the value of $\kappa$ was chosen differently than above, so the states cannot be directly marked in Fig.~\ref{fig:Onset}.

The rightmost column of Fig.~\ref{fig:wf_onset} represents a state which is originally far away from the condensation centers in Fig.~\ref{fig:Onset} (left).
Indeed, this state resists the smaller perturbation rather well.
On the other hand, the most pronounced structural changes at $B=0.005$ and $0.01$ are observed for the states in the first three columns of Fig.~\ref{fig:wf_onset}.
The states in the middle two columns belong directly to the condensation center of strongly interacting levels and the state in the leftmost column is close to it.
One clearly observes the breakdown of the rotational symmetry and a gradual crossover to a trifoliolate form of the wave functions, particularly for the two states in the middle.
While the states with $\Vaverage<0$ are localized more around the minima of the potential, the ones with $\Vaverage>0$ dwell more in the saddle-point regions.
In the fourth row of Fig.~\ref{fig:wf_onset}, which corresponds to the irregular lattice at $B=0.05$, cf. Fig.~\ref{fig:Onset} (c), all four states are already perturbed.
We observe that the form of the rightmost state has been transmitted to the third state in the last row and vice versa, as results from an avoided crossing of both levels.

\begin{figure}[tbp]
		\centering
		\includegraphics[width=\linewidth]{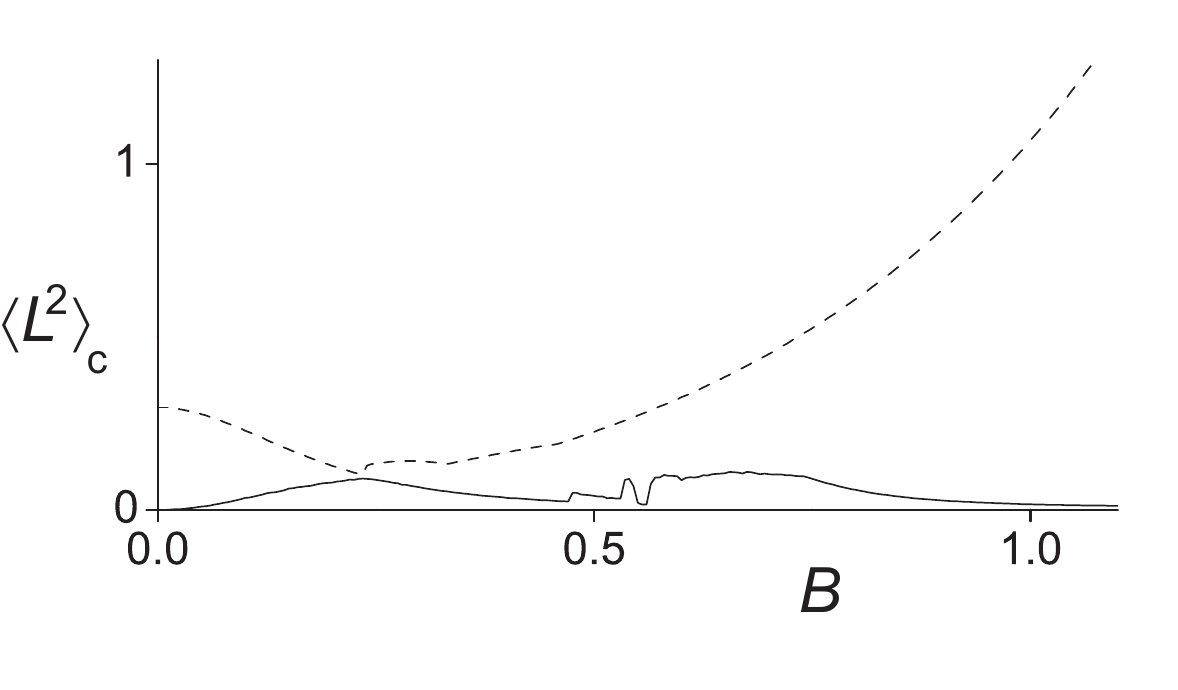}
		\caption{\protect\small 
The lower and higher bounds (solid and dashed lines, respectively) of the classical Peres average $\Laverage_{\rm c}$ for $E=0$.
The bounds almost touch each other at $B=0.24$, where the most chaotic region is located.
Fluctuations of the lower bound near $E=0.5$ are caused by the appearance and disappearance of small unstable tori.
		}		
		\label{fig:L_minmax}
\end{figure}

Finally, it is instructive to look also at the changes of classical Peres invariants with parameter $B$.
We have calculated the lower and upper bounds of the classical average $\Laverage_{\rm c}$ and show the results for $E=0$ in Fig.~\ref{fig:L_minmax}.
As we see, the interval of $\Laverage_{\rm c}$ is contracted almost to a single value in the most chaotic case, $B=0.24$, where the ergodicity is maximal.
Surprisingly, even the small remnants of tori present there have the same value of $\Laverage_{\rm c}$. 
If we step over the most chaotic point, several new tori with higher values of $\Laverage$ appear, which results in a widening of the interval between the bounds.
Note that the narrow ``neck'' in $\Laverage$ at $E=0$ can also be observed in the quantum Peres lattice in Fig.~\ref{fig:Onset}~(e).

\subsection{A quasiregular region}
\label{se:re}

Looking at the form of $\Laverage$ and $\Vaverage$ lattices in row (e) of Fig.~\ref{fig:Onset}, one can notice two characteristic features:
(i) The centroid value of $\Laverage$ exhibits roughly a linear increase with energy $E$.
(ii) The lattice for $\Vaverage$ grows linearly only at low energies, while for higher energies it scatters around a roughly constant average.
These types of dependences in both lattices are qualitatively understandable and remain approximately valid for all increasingly high values of parameter $B$.
Nevertheless, the distribution of regular and irregular parts within the lattices of the above forms exhibits a high degree of variability.

At the first sight, one could expect that the increase of the perturbation strength in the GCM Hamiltonian from $|B|=0$ to $|B|>0$ should lead to a monotonous progression of disorder, as described in the preceding subsection.
Although this scenario is typical for many related systems of type \eqref{hlina}, see e.g. Refs.~\cite{Hal84,Sel84}, it does not apply in the present case.

The GCM is peculiar in two respects.
First, as shown in Ref.~\cite{Cej04}, the classical regular fraction $\freg$ for negative energies converges to unity for asymptotically large values of $B$, when the GCM Hamiltonian can be rescaled to the form $H_{\infty}\equiv T+\beta^4+\beta^3\cos 3\gamma$.
Let us stress that the type of order observed for $E<0$ in the $B\to\infty$ limit is totally different from the $B=0$ case and that in the asymptotic limit the regularity fades away at positive energies.
Second, the competition between regular and chaotic dynamical modes gets surprisingly complex at medium values of $|B|$.
The most important change in this range takes place around $B\approx 0.6$, where extensive regular patterns are established in the Peres lattices at both low and medium energies, see Fig.~\ref{fig:Peres062}.
(Note that the ordered dynamics at very high energies is connected with the dominance of the $\beta^4$ term of the potential \cite{Cej04}.)

The $B\approx 0.6$ quasiregular region was briefly mentioned already in Part~I \cite{I}.
As shown in Ref.~\cite{Mac07}, it is closely related to a so-called \lq\lq arc of regularity\rq\rq\ observed \cite{Alh91} in the parameter space of the interacting boson model (IBM) \cite{ibm}.
Although the IBM is a more sophisticated model of nuclear collectivity that the GCM, it makes use of a similar language (involving quadrupole degrees of freedom) and yields comparable results.
Since its discovery, the IBM arc of regularity has been subject to several analyses \cite{Alh91,Mac07,Cej98,Jol04}.
Some interesting hints have been disclosed, but many questions remained open.

\begin{figure}[tbp]
		\centering
		\includegraphics[width=0.9\linewidth]{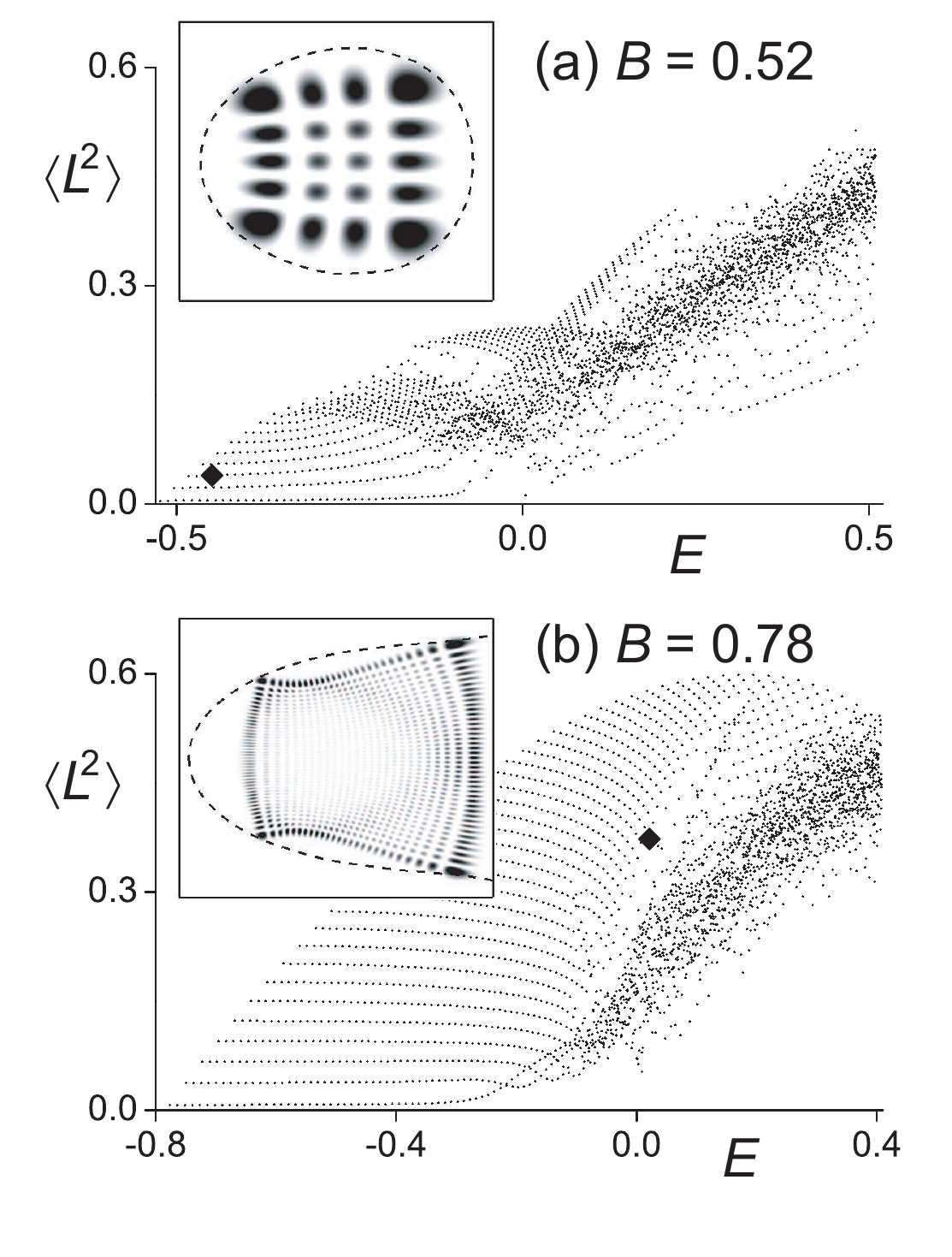}
		\caption{\protect\small 
Peres lattice for $\Laverage$ in 2D even quantization ($\hbar=5\cdot10^{-3}$) at $B=0.52$ (a) and 0.78 (b).
We see that the regular pattern is present before and after the main peak of regularity at $B=0.62$, cf. Fig.~\ref{fig:Peres062}~(a).
Both panels contain the same number of states.
The insets depict selected wave functions (diamonds in the respective lattices) demonstrating the presence of $\beta$ and $\gamma$ vibrations (only the sector around the minimum $\gamma=4\pi/3$ is shown in both cases).
The state in the upper panel ($i=17$) is at $E=-0.449$, the lower one ($i=1292$) at $E=0.021$.
		}		
		\label{fig:arcus}
\end{figure}				

In the GCM case, we observe a phenomenon very similar to the IBM arc.
A comparison of Figs.~\ref{fig:Onset}~(e) and \ref{fig:Peres062}~(a) provides a clear evidence for a large increase of regularity between the two values of $B$.
In fact, the pattern of ordered points, which dominates in the low-energy part of the lattice at $B\approx 0.6$, starts rising already before the maximum of regularity is reached and persists long after it is left.
As an example, we show in Fig.~\ref{fig:arcus} the $\Laverage$ lattices for $B=0.52$ (panel a) and $B=0.78$ (panel b).
It is obvious that the low-energy parts of both lattices exhibit a great deal of similarity with Fig.~\ref{fig:Peres062}~(a).

The mechanism behind the $E<0$ pattern of ordered points visible in all $\Laverage$ lattices at medium and large values of $|B|$ is connected with a competition of two types of vibrations.
To show this, we apply the quadratic-well approximation, valid for $|B|>0$ at low energies above the potential minimum.
It relies on the local use of a 2D oscillator potential
\begin{equation}
V\approx V_0+\frac{k_{\beta}}{2}(\beta-\beta_0)^2+\frac{k_{\gamma}}{2}\beta^2(\gamma-\gamma_0)^2
\end{equation}
where $\beta_0$ and $\gamma_0$ stand for a position of the potential minimum, and $k_{\beta}=\left(\tfrac{\partial^2V}{\partial\beta^2}\right)_0$ and $k_{\gamma}=\left(\tfrac{\partial^2V}{\partial\gamma^2}\right)_0$ for the rigidity of the oscillator in $\beta$ and $\gamma$ directions.

It turns out that the horizontal chains of points with increasing energy, which can be observed in the lattices in Figs.~\ref{fig:Peres062}~(a) and \ref{fig:arcus}~(a,b), correspond to states with a growing number of $\beta$-vibration quanta $n_{\beta}$.
The vertical arrangement of these chains, on the other hand, follows an increasing number of $\gamma$-vibration quanta $n_{\gamma}$.
Such an interplay of vibrational modes in both $\beta$ and $\gamma$ directions represents the basic organization principle for the {\em low-energy\/} part of the $\Laverage$ lattices for $|B|>0$.
Remind that this is essentially different from the $B=0$ situation, when the lattice was determined by vibrational modes in $\beta$ and rotational modes in $\gamma$. 
An example of a $\beta\times\gamma$ vibrational state (its squared wave function) is shown in the inset of Fig.~\ref{fig:arcus}~(a).

A simple calculation shows that at $B=2/3\doteq 0.66$ one gets $k_{\beta}=k_{\gamma}=12$.
We encounter a resonance of the local oscillator frequencies in $\beta$ and $\gamma$ directions, which leads to an additional regularization of the lattice.
Interestingly, due to mutual interactions between levels the degeneracy in a wide interval of energies becomes maximal already at $B=0.62$, where the main peak of regularity takes place.
The resonance is responsible for the \lq\lq condensation\rq\rq\ of the $\Laverage$ lattice at $E<0$ along a nearly horizontal line of multiple points apparent in Fig.~\ref{fig:Peres062}.
Although in Fig.~\ref{fig:Onset} a similar phenomenon was linked to initiating the first seeds of disorder, its role in the present case is rather opposite: it helps to clean up some disarranged parts of the lattice.
Let us stress that the proximity of levels implies rapid structural changes with no immediate relation to chaos.
It can indicate a crossover to chaos as well as emergence of order (imagine the scenario from Sec.~\ref{se:break} played in the reverse direction---with $B$ decreasing to 0).

It needs to be stressed that at $B\approx 0.6$ the patterns emerging in the Peres lattice and in wave functions go far beyond the quadratic-well approximation.
Indeed, as shown in Fig.~\ref{fig:Peres062}, the ordered part of the lattice exceeds to {\em medium energies}, where the approximation deteriorates and even becomes completely invalid (this is certainly so at energy $E_{\rm sad}$ of a saddle point of the potential, where the three regions around the minima merge together and form a single connected area).
Surprisingly, even at $E>0>E_{\rm sad}$ a large fraction of states still keeps the form with well distinguished $\beta$ and $\gamma$ vibrations.
This is exemplified by a selected wave function in the inset of Fig.~\ref{fig:arcus}~(b), where the vibrational pattern remains confined around the potential minimum despite the fact that the energetically accessible domain unifies all three sectors.
Note that peculiar $\beta\times \gamma$ vibrational structures connecting all sectors can be found in even higher eigenstates, cf. Fig.~\ref{fig:wf_chreg}~(c).
One may therefore assume that the observed regular island at $B\approx 0.6$ is due to a fortunate coincidence of resonating $\beta$ and $\gamma$ modes in both $E<0$ and $E\gtrsim 0$ domains.

Qualitatively the same explanation is valid also in the IBM.
There, the degeneracy of $\beta$ and $\gamma$ vibrations was noticed empirically \cite{Jol04} and later supported by theoretical arguments \cite{Mac07}.
A detailed analysis of the $\beta$ and $\gamma$ modes in the IBM framework is in preparation \cite{tobe}.
The present work provides an independent verification of this mechanism in the simpler GCM case.
It also clearly manifests the influence (probably specific for the present form of potential) of the low-energy ordering of states on the spectrum at higher energies, which is significant for the large extension of the regular region.

\section{Conclusions}\label{sec:Conclusions}

In this paper, we have continued and exceeded the work presented in Part~I \cite{I}, whose main purpose was to test the Bohigas conjecture for different quantization schemes under the condition of a strong variability of chaotic measures with energy.
We have revitalized an almost 35 years old method by Peres \cite{Per84} and showed its great potential in the field of quantum chaos and even beyond.

Peres lattices provide an excellent viewpoint to the landscape around the border between classical and quantum physics.
This is so especially for systems with two degrees of freedom whose lattices can be drawn as two-dimensional diagrams, in analogy with planar Poincar{\'e} sections of such systems.
If applied within the domain of quantum chaos, Peres' method enables one to distinguish regular and chaotic behaviors on the level of individual states or subsets of states within the same energy interval.
This is in contrast to traditional methods based on spectral statistics that assign the same degree of chaos to all levels within the same interval.

Quite naturally, there are some limitations of the method.
As seen, regular and irregular parts of the lattice can in some cases be superimposed on each other, which hinders their correct resolution. 
In particular, the distinction of chaotic states may be ambiguous since a superposition of two or more regular patterns may seem irregular.
Nevertheless, we showed that this problem can in principle be bypassed by constructing more lattices with different Peres operators.
Their optimal choice, which unavoidably depends on the concrete system under consideration, should be subject to further study.
Although the Peres' method does not directly yield a calculable measure of quantum chaos, it represents an important indicator providing new insights into the origin of chaotic behavior.

However, our intention in this paper was to go even beyond the scope of quantum chaos, demonstrating that Peres lattices represent an extremely efficient and economic tool for studying significant features in large ensembles of eigenstates across the spectrum.
Relevant properties of the wave functions can be read off from the expectation values of suitably chosen Peres operators.
Instead of analyzing each individual eigenstate and its wave function, one may look at the associated Peres lattice where the desired information is contained in a synoptical way.
As an example, we were able to closely follow the breakdown of integrability of the system and the rise of a new type of order.
We believe that the results presented here may encourage similar studies in other systems.

The present work completes our long-term effort to map chaotic properties of the geometric collective model of nuclear physics \cite{I,Cej04}.
A great advantage of the geometric model (and also of the related interacting boson model) is the apparent conceptual simplicity encoding strikingly rich complexity of dynamics.
Let us note that the above simplified models capture the main phenomenological features of nuclear collectivity which are presently beside a fully microscopic description.
The study of disordered collective dynamics within these models may be considered as an attack to the problem of chaos in many-body systems from the direction perpendicular to the mean-field approach.

An interactive survey of our main results can be found at the website \cite{www}.


\acknowledgments
We would like to thank A. Frank, A. Leviatan, and M. Macek for inspiring discussions.
This work was supported by Czech Science Foundation (grant no. 202/06/0363) and by the Czech Ministry of Education (contracts nos. 0021620859 and  LA 314).


\end{document}